\documentstyle[aps,prl,epsf,twocolumn]{revtex}
\begin{document}
\draft

\wideabs{
\title{Baryon phase-space density in heavy-ion collisions}

\author{Fuqiang Wang and N. Xu}
\address{Nuclear Science Division, Lawrence Berkeley National Laboratory,
	Berkeley, CA 94720, USA}

\maketitle

\begin{abstract}
The baryon phase-space density at mid-rapidity from central heavy-ion
collisions is estimated from proton spectra with interferometry and
deuteron coalescence measurements. 
It is found that the mid-rapidity phase-space density of baryons is 
significantly lower at the SPS than the AGS, while those of total 
particles (pion + baryon) are comparable. 
Thermal and chemical equilibrium model calculations tend to 
over-estimate the phase-space densities at both energies.
\end{abstract}

}

Baryon phase-space density affects the dynamical evolution of 
heavy-ion collisions from the initial stage to final freeze-out.
It is one of the essential ingredients of thermal and chemical equilibrium 
models often used in studying heavy-ion collisions~\cite{Bra96,Bec98}.
The studies of baryon and meson phase-space densities may provide
information on particle freeze-out conditions and entropy production
in heavy-ion collisions~\cite{Sie79,Nag86}.
Experimentally, phase-space densities can be estimated from
particle yields and source sizes inferred from two-particle 
interferometry \cite{Bertsch,Jac99,Panitkin} and, in the case of nucleons, 
from deuteron coalescence \cite{Sie79,Nag86,Cse86}.
The values of the phase-space densities may be then compared to 
Bose-Einstein or Fermi-Dirac statistics to assess the degree to 
which thermal and chemical equilibria are established 
\cite{Bar97,Fer99,Tom99}.
One unique advantage of such comparisons is that they do not require
the extrapolation of data to all of phase-space~\cite{Cle99,Voloshin}.

The freeze-out particle spatial density is also of interest.
However, it is known that, in high energy heavy-ion collisions, the 
freeze-out space point of a particle is correlated with its momentum.
Because of this space-momentum correlation, the size parameters 
extracted from two-particle interferometry at low relative momentum 
do not reflect the full size of the particle-emitting source~\cite{Vos94}.
Hence, one cannot readily deduce the particle spatial density from 
the size parameters and particle yields which is from the entire source.

There have been several studies of the freeze-out 
phase-space density of pions \cite{Bertsch,Bar97,Fer99,Tom99}.
In this paper, we study the freeze-out phase-space density of baryons.
We first estimate the baryon phase-space density at mid-rapidity in heavy 
ion collisions at the AGS and SPS using two-proton interferometry data. 
We then investigate the ratio of deuteron to proton invariant cross-sections
as a means to extract the baryon phase-space density~\cite{Sie79,Cse86}.
Finally, we compare our estimates to thermal and chemical equilibrium
model calculations.

We start from the definition of particle phase-space density,
\begin{equation}
f({\bf p},{\bf x}) \equiv \frac {dN} {d{\bf p}d{\bf x}},
\label{eq:f(p,x)definition}
\end{equation}
which is Lorentz invariant.
Consider identical particles in a small momentum cell,
(${\bf p}, {\bf p}+d{\bf p}$), 
which occupy a spatial volume $V({\bf p})$ in the particle rest frame. 
(We specify $V({\bf p})$ in the particle rest frame because available 
two-proton correlation measurements~\cite{NA49_2p,E877_2p} are done 
in such frame.)
If these particles are uniformly distributed, 
then the spatial averaged phase-space density is
\begin{equation}
\langle f ({\bf p}) \rangle = 
\frac{E}{m} \frac {dN/d{\bf p}} {V({\bf p})},
\label{eq:f(p)}
\end{equation}
where $m$ is the particle rest mass, and $E/m$ is due to the Lorentz boost.

Instead of a uniform distribution in space, if we take a Gaussian form 
for the particle density profile (as often used to extract source size 
parameters from two-particle interferometry~\cite{Boa90,Gel92}), 
\begin{equation}
f({\bf p},{\bf x}) = \frac{E}{m} 
\frac {dN/d{\bf p}} {\left[ \sqrt{2\pi} R_G({\bf p}) \right]^3} 
e^{-\left[ {\bf x}-{\bf x_0({\bf p})} \right]^2 / 2R^2_G({\bf p})},
\label{eq:f(p,x)gauss}
\end{equation}
then Eq.~(\ref{eq:f(p)}) becomes
\begin{equation}
\langle f ({\bf p}) \rangle = 
\frac {\int f^2({\bf p},{\bf x}) d{\bf x}}
      {\int f  ({\bf p},{\bf x}) d{\bf x}} = \frac{E}{m} 
\frac {dN/d{\bf p}} {\left[ 2\sqrt{\pi}R_G({\bf p}) \right]^3}.
\label{eq:f(p)gauss}
\end{equation}
$R_G$ is a measure of the relative separation 
of two particles close in momentum in the pair rest frame 
when the later particle freezes out~\cite{Vol97}.
In the above equations, however, it has been assumed that 
the particles freeze-out at an instant time, 
and the phase-space density is calculated at that time. 
When the freeze-out process has a finite time duration,
the phase-space density should be considered as a 
``time-averaged'' quantity.

Without knowing the exact form of the space-momentum correlation, 
${\bf x_0({\bf p})}$, one cannot perform the integration over {\bf p}. 
Hence, one cannot obtain the spatial density.
However, the operation can be carried out over the spatial 
coordinate {\bf x} leading to Eq.~(\ref{eq:f(p)gauss}).
In the following discussions, we will use the Gaussian density profile
and denote $V({\bf p}) = \left[ 2\sqrt{\pi}R_G({\bf p}) \right]^3$.

The averaged phase-space density weighted 
by particle number density is given as
\begin{equation}
\langle f \rangle = \frac{\int \langle f({\bf p}) \rangle dN}{\int dN} 
                  = \frac{
\int \frac{1}{mV({\bf p})} \left( E\frac{dN}{d{\bf p}} \right)^2 
\frac{d{\bf p}}{E} }
{ \int \left( E\frac{dN}{d{\bf p}} \right) \frac{d{\bf p}}{E} }.
\label{eq:f}
\end{equation}
We are mainly interested in the baryon phase-space density.
Since the transverse momentum ($p_T$) dependence of $R_G$ for 
mid-rapidity proton source has not been measured, we assume an
$p_T$ independent volume $V(y,p_T)=V(y)$ in the following discussions.
Taking an exponential function in $m_T=\sqrt{m^2+p^2_T}$ 
for the invariant cross-section 
$E\frac{dN}{d{\bf p}} = \frac {dN/dy} {2\pi T (T+m)} e^{-(m_T-m)/T}$,
we obtain 
\begin{eqnarray}
\langle f(y) \rangle 
& = &	\frac {dN/dy} {2\pi m T^2 (T+m)^2} \int_m^{\infty} 
	\frac { e^{-2(m_T-m)/T} } { V(y, p_T) } m_T dm_T \nonumber \\
& = &	\frac {T+2m} {8\pi m T (T+m)^2} \frac {dN/dy} {V(y)}.
\label{eq:f(ynn)}
\end{eqnarray}

Bertsch obtained similar results for mesons from two-particle 
correlation formula~\cite{Bertsch}, assuming an exponential 
transverse distribution in $p_T$.
Our results above are valid for both mesons and baryons, 
and are consistent with Bertsch's if same assumptions are made.

Using Eq.~(\ref{eq:f(ynn)}) and available data 
\cite{NA49_2p,E866ppi,AGS2pi,NA49ppi,NA44ppi,NA44p,NA49_2pi,NA44_2pi}, 
the average phase-space densities are estimated for freeze-out pions,
protons and baryons at mid-rapidity in central Au+Au or Pb+Pb 
collisions at the AGS and SPS.
The results are listed in Table~\ref{tab:data}. 
The statistical and systematic errors on the phase-space densities 
are estimated from the corresponding errors on $dN/dy$, $T$, and $R_G$.
In estimating the baryon phase-space density, it is assumed that 
the source sizes of other baryons are the same as proton's.

The source size parameters $R_G$ 
are estimated from pion and proton interferometry measurements in 
invariant relative momentum \cite{NA49_2p,AGS2pi,NA49_2pi,NA44_2pi}.
For pions, $R_G({\bf p})$ is observed to decrease with $p_T$ 
\cite{NA49_2pi,NA44_2pi,Bek95,Kai97} due to the space-momentum correlation.
However, we have only used the low $p_T$ pion source size in our estimation.
This introduces an underestimate of the pion phase-space density,
for instance, by about 25\% for the SPS data.
For the mid-rapidity proton source size at the AGS,
we have used the two-proton correlation measurement at 
forward rapidity~\cite{E877_2p} and estimated as 
$R_G = 3.5\pm 0.5$~(syst.)~fm such that the low end of the 
systematic error equals to the initial nucleus size $R_G=3.0$~fm.
This estimate is consistent with a hypothetical $1/\sqrt{m_T}$ scaling
from pion interferometry measurements.
The errors introduced by assuming a $p_T$ independent volume are
not included in the systematic errors quoted in Table~\ref{tab:data}.

It is known that the pion and proton transverse distributions 
cannot be satisfactorily described by a single $m_T$ exponential 
\cite{E866ppi,NA49ppi,NA44ppi,NA44p}. 
In order to check the errors, we used the available data to 
calculate the phase-space density by Eq.~(\ref{eq:f}) and 
the results are consistent with the above estimates within 15\%. 
This is also reflected in the systematic errors on the inverse 
slope parameter ($T$) quoted in Table~\ref{tab:data}. 
In fact, the values of the phase-space density extracted from 
this work are consistent with those obtained in \cite{Bar97,Fer99,Tom99}.

As seen in Table~\ref{tab:data}, the freeze-out pion phase-space
densities are similar ($\sim$0.1) for both AGS 
($\sqrt{s}\approx 5$ AGeV) and SPS ($\sqrt{s}\approx 20$ AGeV) energies. 
But the proton(baryon) phase-space density is significantly 
lower than the pion's at both beam energies. 
On top of that, the mid-rapidity baryon phase-space density 
decreases as the beam energy increases.
The similarity of the total phase-space density (pion + baryon) 
at freeze-out perhaps indicates that, in high energy collisions, 
the freeze-out process occurs at a fixed condition 
(local temperature and density) which is determined by the free 
cross sections as argued by Pomeranchuk in 1951~\cite{pomeranchuk51}. 
Tevatron data seem to indicate a constant particle density 
in $p\bar{p}$ annihilations~\cite{Ale93}.
Note that at the AGS energy, pion to proton ratio is about one 
so the baryon density is the relevant control parameter for 
freeze-out~\cite{heinz}.
At the SPS energy, the pion to proton ratio is about six 
making mesons the dominant player at freeze-out. 
This will be certainly true in the upcoming RHIC collisions
at $\sqrt{s} = 200$ AGeV.

We have assumed constant $V({\bf p})$ 
in estimating $\langle f \rangle$ above.
This is because, as mentioned before, 
a $p_T$ dependent two-proton correlation measurement is not 
presently available in high energy nucleus-nucleus collisions 
due to the lack of statistics.
Such measurement, however, may be obtained from deuteron coalescence 
\cite{Mek_prc,Mek_npa,Nag94,Llo95,Sch99,Mon99,Pol99}.

In the coalescence model with a Gaussian density profile,
$R^3_G({\bf p}_{\rm p}) = \frac{3}{4} \pi^{3/2} 
\left( \frac{E_{\rm p}}{m_{\rm p}} \frac{dN_{\rm p}}{d{\bf p}_{\rm p}} 
\right)^2 \left/ 
\left( \frac{E_{\rm d}}{m_{\rm d}} \frac{dN_{\rm d}}{d{\bf p}_{\rm d}} 
\right)_{{\bf p}_{\rm d}=2{\bf p}_{\rm p}} \right.$,
where the factor $3/4$ comes from spin consideration, 
and subscripts `p' and `d' denote proton and deuteron, respectively.
From Eq.~(\ref{eq:f(p)gauss}), we obtain
\begin{equation}
\langle f ({\bf p_{\rm p}}) \rangle = \frac{4}{3(2\pi)^3} 
\left( \frac{E_{\rm d}}{m_{\rm d}} \frac{dN_{\rm d}}{d{\bf p}_{\rm d}} 
\right)_{{\bf p}_{\rm d}=2{\bf p}_{\rm p}} \left/
\left( \frac{E_{\rm p}}{m_{\rm p}} \frac{dN_{\rm p}}{d{\bf p}_{\rm p}} 
\right) \right.,
\label{eq:f(p)deuteron}
\end{equation}
where the neutron and proton differential cross sections 
are assumed to be identical~\footnote[2]{
Various models~\cite{Sor95,Wer93} show that the neutron to proton 
$dN/dy$ ratio at mid-rapidity in Au+Au (AGS) and Pb+Pb (SPS) 
central collisions is about 1.05.}.
The average proton phase-space density [Eq.~(\ref{eq:f})]
can be approximated as
$\langle f \rangle = \frac{1}{6(2\pi)^3}
\int dN_{\rm d} \left/ \int dN_{\rm p} \right.$.
Therefore, the $p_T$ averaged proton phase-space density is
\begin{equation}
\langle f(y) \rangle = \frac{1}{6(2\pi)^3} 
\frac {(dN/dy)_{\rm d}} {(dN/dy)_{\rm p}},
\label{eq:f(y)deuteron}
\end{equation}
and that averaged over the whole phase-space is
\begin{equation}
\langle f \rangle = \frac{1}{6(2\pi)^3} \frac {N_{\rm d}} {N_{\rm p}},
\label{eq:f_deuteron}
\end{equation}
as pointed out long ago by several authors~\cite{Sie79,Cse86}.

Using Eq.~(\ref{eq:f(y)deuteron}) and available deuteron data 
\cite{AGS2pi,Ahl99,Han99,Mur99}, 
we estimate the proton phase-space density at the AGS and SPS 
to be $\langle f \rangle (2\pi)^3 = 0.015 \pm 0.001$~(stat.)
and $0.0029 \pm 0.0004$~(stat.), respectively.
These results are in a good agreement with those obtained 
using two-proton interferometry data (Table~\ref{tab:data}).

Now we compare our results to calculations by 
local thermal and chemical equilibrium models. 
In these models, temperature ($T_0$) and baryon chemical 
potential ($\mu_B$) at chemical freeze-out are extracted 
from particle ratios \cite{Bra96,Bec98,Bra95,Bra99,Bra_private}; 
temperature at kinetic freeze-out is extracted from
particle spectra with certain assumptions about collective 
flow~\cite{NA49_2pi,heinz}.
These results as obtained from AGS and SPS data 
are reproduced in Table~\ref{tab:thermal}. 

In local thermal and chemical equilibrium models, 
the phase-space density is 
$\langle f(E) \rangle = 
(2\pi)^{-3} \left/ \left[ e^{E/T_0} - 1 \right] \right.$ 
for pion and
$\langle f(E) \rangle = 
2(2\pi)^{-3} \left/ \left[ e^{(E-\mu_B)/T_0} + 1 \right] \right.$
for proton, respectively.
Here $E$ is the particle energy, $T_0$ the temperature, 
and $\mu_B$ the baryon chemical potential.
The average particle phase-space densities at mid-rapidity, 
$\langle f \rangle = 
\int \langle f(E) \rangle^2 dp^2_T \left/
\int \langle f(E) \rangle   dp^2_T \right.$, 
are calculated separately for chemical and kinetic freeze-out.
The results are listed in Table~\ref{tab:thermal}.
The phase-space densities at chemical freeze-out are higher 
by a factor of $2.7 \pm 1.2$ (AGS) and $4.0 \pm 1.3$ (SPS) 
than estimated from experimental data (Table~\ref{tab:data}).
Likewise, the phase-space densities calculated for kinetic freeze-out 
are higher by a factor of $1.5 \pm 0.7$ (AGS) and $2.2 \pm 1.2$ (SPS).

One important assumption in local thermal and chemical equilibrium 
models is that all particles freeze-out at the same time, and they occupy 
the same volume so that the volume is divided out in particle ratios.
From this assumption and the fact that particle numbers stay constant
after chemical freeze-out, an extrapolation has been made for the
chemical potential from chemical to kinetic freeze-out. 
This is equivalent to the assumption that the system undergoes isentropic
expansion after chemical freeze-out~\cite{Cle99,Hei99}~\footnote[3]{ 
In this procedure, it is explicitly assumed that local thermal and
chemical equilibrium is always maintained for pions and protons 
during expansion of the collision system. These pions include 
resonance decay products.}. 
After chemical freeze-out, the system expands and the temperature 
decreases. 
This leads to a factor of 3 increase in the volume at both the AGS 
and SPS energies. 
Taking the average radial flow velocity as 
$v_{\rm f}\approx 0.5c$~\cite{NA44ppi}, 
one may estimate the corresponding expansion time: 
$(\sqrt[3]{3}-1)R_A/v_{\rm f}\sim 6$~fm/$c$ with $R_A = 1.2A^{1/3}$.

In summary, we have studied the mid-rapidity 
baryon phase-space density at freeze-out through 
(1) proton yield and two-proton interferometry measurements, and 
(2) the ratio of deuteron to proton differential cross-sections. 
For heavy-ion collisions at both the AGS and SPS, the estimated 
phase-space density of baryons is significantly lower than of pions. 
The baryon phase-space density at the AGS is about three times that 
at the SPS, while the values for the pion phase-space density are 
similar at both energies. 
The thermal and chemical equilibrium model calculations are higher
than experimental data at both beam energies.

Note that the present analyses suffer from large systematic 
uncertainties. 
In order to obtain more precise results for baryons, the proton 
interferometry or deuteron coalescence data as a function of 
the transverse momentum are needed.
In addition, we only studied the particle phase-space density 
at mid-rapidity.
It will be interesting to repeat the study at projectile/target 
rapidities and compare with thermal model predictions.

\acknowledgements{
We are grateful to Dr. S. Voloshin for stimulating discussions.
Discussions with Dr. D. Brown, Dr. S. Panitkin and Dr. S. Pratt 
are greatly acknowledged. 
We thank G. Cooper and Dr. U. Heinz for critical readings 
of the manuscript.
This work was supported by the U.S. Department of Energy 
under contract DE-AC03-76SF00098.}

\onecolumn

\begin{table}
\caption{
The average phase-space density $\langle f \rangle$ estimated
for freeze-out pions, protons and baryons at mid-rapidity in 
central Au+Au or Pb+Pb collisions at the AGS (upper section) 
and SPS (lower section).
Input parameters are mid-rapidity $dN/dy$, inverse slope $T$, 
and Gaussian source size $R_G$ in the pair rest frame. 
Unless specified, for each quantity the first error 
is statistical and the second error is systematic.}
\label{tab:data}
\begin{tabular}{cllll}
particle& $dN/dy$ 
	& $T$ (MeV) 
	& $R_G$ (fm) 
	& $\langle f \rangle (2\pi)^3 $ 
\\ \hline
$\pi^+ \approx \pi^-$
	& $62 \pm 2 \pm 8$	\cite{E866ppi} 
	& $170 \pm 2 \pm 10$	\cite{E866ppi} 
	& $6.2 \pm 0.4 \pm 0.4$	\cite{AGS2pi} 
	& $0.087 \pm 0.017 \pm 0.022$
\\
proton 	& $61 \pm 1 \pm 8$	\cite{E866ppi} 
	& $270 \pm 3 \pm 10$	\cite{E866ppi} 
	& $3.5 \pm 0.5$ (syst.)	
	& $0.014 \pm 0.006$ (syst.)
\\
baryon 	& $140 \pm 2 \pm 20$	\cite{E866ppi} 
	& $\prime\prime$
	& $\prime\prime$
	& $0.032 \pm 0.015$ (syst.)
\\ \hline
$\pi^+ \approx \pi^-$
	& $180 \pm 10 \pm 20$	\cite{NA49ppi,NA44ppi} 
	& $206 \pm 2 \pm 20$ 	\cite{NA49ppi,NA44ppi} 
	& $7.4 \pm 0.4$ (stat.+syst.)	\cite{NA49_2pi,NA44_2pi} 
	& $0.106 \pm 0.028$ (stat.+syst.)
\\
proton	& $27 \pm 4 \pm 5$	\cite{NA49ppi,NA44ppi,NA44p} %
	& $285 \pm 5 \pm 30$ 	\cite{NA49ppi,NA44ppi,NA44p} 
	& $3.85 \pm 0.15 ^{+0.60}_{-0.25}$	\cite{NA49_2p}
	& $0.0043 \pm 0.0008 ^{+0.0013}_{-0.0023}$ 
\\
baryon 	& $68 \pm 8 \pm 10$	\cite{NA49ppi,NA44ppi,NA44p} 
	& $\prime\prime$
	& $\prime\prime$
	& $0.0109 \pm 0.0018 ^{+0.0031}_{-0.0056}$
\\
\end{tabular}
\end{table}

\begin{table}
\caption{
The average mid-rapidity phase-space densities for pions 
and protons at chemical ($\langle f \rangle_{\rm ch}$) and 
kinetic ($\langle f \rangle_{\rm kin}$) freeze-out from 
the thermal and chemical equilibrium model. 
Input parameters are temperatures ($T_0$ in MeV) at chemical 
and kinetic freeze-out, and baryon chemical potential ($\mu_B$ in MeV) 
at chemical freeze-out extracted from central Au+Au or Pb+Pb 
experimental data at the AGS (upper section) and SPS (lower section). 
All errors are statistical only. 
See text for estimation of $\mu_B$ at kinetic freeze-out.}
\label{tab:thermal}
\begin{tabular}{ll||cll}
		Chemical freeze-out			& 
		Kinetic  freeze-out \hspace{0.4in}	&
		particle			& 
		$\langle f \rangle_{\rm ch}  (2\pi)^3 $	&
		$\langle f \rangle_{\rm kin} (2\pi)^3 $
		\\ \hline
		$T_0 = 125^{+3}_{-6}$ \cite{Bra95,Bra99,Bra_private}	&
		$T_0 = 92 \pm 2$ \cite{heinz}	&
		$\pi^+ \approx \pi^-$	&
$0.172^{+0.006}_{-0.011}$	&
$0.107 \pm 0.004$
				\\
		$\mu_B=540 \pm 7$ \cite{Bra95,Bra99,Bra_private}	&
$\mu_B=590 \pm 5$
		& proton	& 
$0.038^{+0.004}_{-0.006}$	& 
$0.021 \pm 0.002$
		\\ \hline
		$T_0 = 168 \pm 2.4$ \cite{Bra99,Bra_private}	& 
		$T_0 = 120 \pm 12 $ \cite{NA49_2pi}
		& $\pi^+ \approx \pi^-$	& 
$0.249 \pm 0.004$		& 
$0.162 \pm 0.023$
		\\
		$\mu_B = 266 \pm 5$ \cite{Bra99,Bra_private} &
$\mu_B = 389 \pm 4$
		& proton	&
$0.017 \pm 0.001$		& 
$0.0096 \pm 0.0043$
\end{tabular}
\end{table}

\end{document}